\documentstyle[psfig,prb,twocolumn,amsmath,aps]{revtex}

\begin{document}
\draft
\pagestyle{empty}
\sloppy

\title{On the dynamics of coupled S =1/2 antiferromagnetic zig-zag chains}
\author{M. M\"uller and H.-J. Mikeska}
\address{Institut~f\"ur~Theoretische~Physik, Universit\"at~Hannover, 
30167~Hannover, Germany}

\date{May 19, 2000}

\maketitle 
\begin{abstract} 
We investigate the elementary excitations of quasi one-dimensional
$S=\frac{1}{2}$ systems built up from zig-zag chains with general
isotropic exchange constants, using exact (Lanczos) diagonalization
for 24 spins and series expansions starting from the decoupled dimer
limit. For the ideal one-dimensional zig-zag chain we discuss the
systematic variation of the basic (magnon) triplet excitation with
general exchange parameters and in particular the presence of
practically flat dispersions in certain regions of phase space. We
extend the dimer expansion in order to include the effects of 3D
interactions on the spectra of weakly interacting zig-zag chains. In
an application to $\rm KCuCl_3$ we show that this approach allows to
determine the exchange interactions between individual pairs of spins
from the spectra as determined in recent neutron scattering
experiments.
\end{abstract} 
\vspace*{0.6cm}

\noindent

\vskip 0.5cm 

\section{Introduction}

Spin systems consisting of chain- or ladderlike structures as basic
building blocks have recently attracted much attention. These systems
are of interest on the one hand as one-dimensional (1D) model systems
allowing to study quantum phase transitions related to the existence
of a spin gap and their dependence on the exchange parameters
\cite{DagR96,BMN96}; on the other hand they describe an increasing
number of real materials when an additional (small) exchange coupling
in the remaining two dimensions is introduced \cite{Dag99}. A material
of particular recent experimental interest is $\rm KCuCl_3$
\cite{KTTS98,CHF99}.
\begin{figure}
\mbox{\psfig{figure=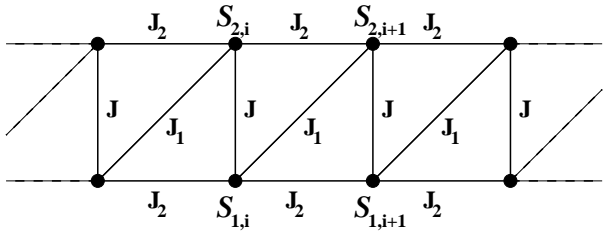,width=80mm,angle=0.}}
  \vskip 2mm\nopagebreak
\caption{\label{fig:leitmod} Zig-zag chain as defined by (~\ref{eq:hamilton})}
\end{figure}
We have performed an investigation of the dynamics of such systems
with a twofold aim: (i) We discuss the spectrum $\omega(q_x)$ of the
low-lying triplet excitations in the ideal 1D system over a wide range
of exchange parameters using both series expansions and exact
diagonalization, in order to determine the range of applicability of
the series expansion approach and to study the validity of using an
effective interaction between dimers. We find and discuss in
particular a regime in phase space with extremely small dispersion and
a minimum of $\omega(q_x)$ at finite wavevector $0 < q_x < \pi$.  (ii)
We extend the dimer expansion to include 3D couplings and apply this
method in particular to a discussion of the dynamics of the quasi
zigzag-ladder material KCuCl$_3$ in terms of microscopic exchange
parameters.

Of particular interest as an 1D building block for this type of materials
is the $S=\frac 1 2$ zig-zag chain, as shown in Fig.~\ref{fig:leitmod} and defined
by the following hamiltonian:

\begin{eqnarray}\label{eq:hamilton}
H = \sum_{i=1}^L J \vec S_{1,i} \vec S_{2,i} 
     &+& J_1 \vec S_{1,i} \vec S_{2,i+1} \nonumber \\ 
     &+& J_2 \left( \vec S_{1,i} \vec S_{1,i+1} 
                          + \vec S_{2,i} \vec S_{2,i+1} \right). 
\end{eqnarray}

On the theoretical side, this generic 1D model interpolates between a
number of seemingly different limiting cases: It is an alternative way
to formulate the hamiltonian for the generalized $S=\frac 1 2$ spin
ladder generalized to include one diagonal interaction or equivalently
the $S=\frac 1 2$ chain with nearest neighbour (NN) alternating
exchange and next nearest neighbour exchange (we use the shorthand
NNNA-chain in the following). It thus covers the well-known limiting
models of the isotropic $S=\frac 1 2$ Heisenberg chain (HAF, $J =
J_1, J_2=0$), the standard antiferromagnetic $S=\frac 1 2$ ladder
($J_1 = 0, J = J_2 > 0$), the weakly interacting dimer chain ($J_1,
J_2 \ll J$) and the $S=1$ antiferromagnetic (Haldane) chain ($J_1 \to
-\infty, J+2J_2 > 0$). It can alternatively be considered as a two
legged spin ladder with rung coupling $J$, leg coupling $J_2$ and
additional diagonal coupling $J_1$.

The theoretical interest in the dynamics of the NNNA-chain goes back
to the work of Shastry and Sutherland \cite{SS81}, who identified the
elementary excitations without alternation as free particles
(spinons), which may become bound. A variational approach to the
excitations of the NNNA-chain based on this concept \cite{BKMN98} has
recently been shown to cover qualitatively the transition from free
spinons to the Haldane triplet. 

In a first approximation real materials are often considered as
examples of 1D chains with a hamiltonian as given in Eq.~(1); then
they realize different points in the phase diagram spanned by the
interaction constants $J_1, J_2$ and illustrate the interest to
describe systematically the variation of static and dynamic properties
with the parameters $J_1, J_2$. Static properties, such as
susceptibility and specific heat, however, have turned out to be
rather insensitive to the details of the microscopic hamiltonian
\cite{NO98} and, in the case of $\rm (VO)_2P_2O_7$ 
\cite{GarNT97,WeiBF98,UhrN98} have even not been able to reveal the 
basic interactions as two- instead of one-dimensional. 

Thus for a description of real materials a systematic microscopic
treatment of the dynamics is of particular importance. In section II
we present a systematic overview of the dynamics of the 1D system,
discussing both general properties as well as comparing results from
exact diagonalization to results from series expansions.  In section
III we use the series expansion approach to calculate the low-lying
excitations in a 3D material with the structure of $\rm KCuCl_3$ in
terms of the microscopic hamiltonian. This will allow us to go beyond
the determination of effective dimer exchange parameters in recent
work \cite{KTTS98,CHF99} and to determine the microscopic exchange
parameters. A summary will be given in the concluding section IV.

\section{Elementary excitations of the 1D zig-zag chain} 

We start with a short summary of the symmetries of the 1D system (the
chain direction is denoted as $x-$axis): Translational symmetry is
described by the wavevector $q_x$ defined in a Brillouin zone $-\frac{\pi} a 
< q_x < + \frac{\pi} a$; the unit cell of length $a$ contains
two spins, or equivalently one dimer (one singlet in the limit
$J_1=J_2=0$). We thus use the conventional notation for ladders: $a$
is the distance between rungs, whereas the distance between spins in
the NNNA-chain picture is $a/2$.  We thus expect two basic excitations
per unit cell. We will use units $a=1$ in the following.  Excitation
frequencies at wavevectors $q_x$ and $-q_x$ are equal owing to
reflection symmetry along the chain.

For special points in the phase diagram additional symmetries exist:

Without alternation ($J_1 = J$) it is natural to use a unit cell of
length $\tilde{a} = \frac{1}{2}a$ containing only one spin. Our
Brillouin zone is half of this Brillouin zone of the uniform chain and
the excitations of the conventional spin chain will appear folded
back to our smaller Brillouin zone.
  
For the ladder symmetry ($J_1 = 0$) there exists a quantum number
parity, $P$, resulting from the interchange of the two legs and we can
classify states as positive or negative under this reflection.  Each
dimer in the singlet (triplet) state contributes a factor of $-1 (+1)$
to this parity. An alternative notation introduces the component
$q_{\perp}$ with values 0 (corresponding to $P = +1$) and $\pi$
(corresponding to $P = -1$).

The ground state is a singlet in the whole phase plane and the lowest
excited state is generally a triplet. The ground states for the
ladder symmetry ($J_1=0$) have parity $P=+1$ for $L$ even.

We have studied the dispersion $\omega(q_x)$ of the basic
triplet excitation for a typical variety of paths in the $J_1-J_2$
parameter space by two methods:
(i) By exact numerical diagonalization, using the Lanczos algorithm,
we have calculated $\omega(q_x)$ for the lowest excited
states (between 2 and 4 states) for 24 spins, i.e. for 7 different
values of wavevector $q_x$.
(ii) We have performed series expansions around the dimer point, $J_1
= J_2 = 0$, up to third order analytically and up to 10th order after
implementation of the cluster algorithm \cite{GeSiHu90,Gel96} on an
Alpha work station. Thus we have obtained the ground state energy
$E_0$ and an effective Hamiltonian which can be diagonalized by a
Fourier transformation. Finally we get the dispersion relation for the
lowest excited state expressed as series $\sum_n a_n\cos(nq_x)$.  

In the follwoing we present a number of results for the 1D zig-zag
chain which prepare the stage for the first application of the method
to a nontrivial 3D system in section III and also add some new aspects
to the large number of previous studies on the 1D system defined by
eq. (1) in recent years. To give a short review of existing work we
mention first that the dimer series expansion approach started when
the work of Brooks Harris \cite{Bro73} was revived by Uhrig
\cite{Uhr97} in the context of $\rm CuGeO_3$. The expansion for the
triplet dispersion was extended to high orders recently by Oitmaa et
al \cite{OitSW96} for ladders to $8^{\rm th}$ order, by Barnes et al
\cite{BarRT98} for the Heisenberg alternating chain to $9^{\rm th}$
order and by Singh et al \cite{SinZ99} for the disorder line to
$23^{\rm th}$ order (using a special symmetry on this line). The model
of Eq. (1) was also treated by alternative methods as random phase
approximation \cite{UhrS96}, Br\"uckner theory for the equivalent
dilute Bose gas \cite{KotSE99,SheKS99}, exact diagonalization
\cite{BouKJ98}, and continued fraction expansion based on ED results
\cite{MorHT98} and DMRG \cite{PatCS97}. From these studies a rather
complete picture of the low-energy dynamics of the 1D zigzag chain has
emerged. In this section we supplement this picture by the following
two remarks.

We start from the neighbourhood of the dimer point where an expansion
in $J_1, J_2$ to low orders is sufficient. Up to third
order the following result for the dispersion is obtained (as given in
ref.~\onlinecite{Uhr97}, frequency and exchange constants are measured
in units of the intradimer exchange $J$ from now on):

\begin{eqnarray}
\omega(q_x) &=& 1 - \frac{J_1^2} 4 \left(1+J_2\right) + \frac 3 8\!\left(J_2 - \frac{J_1} 2 \right)^2\!\left(2+J_2- \frac{J_1} 2 \right) \nonumber \\
&+&\left[J_2-\frac{J_1} 2 - \frac{J_1^2} 4 \left(1+J_2\right) - \frac 1 4 \left(J_2 - \frac{J_1} 2\right)^3\right]\!\cos q_x
    \nonumber \\     
  &-& \frac 1 4 \left(J_2 - \frac{J_1}{2}\right)^2\left(1+J_2+\frac{J_1} 2 \right)\!\cos 2q_x \nonumber \\
  &+& \frac 1 8 \left(J_2 - \frac{J_1}{2}\right)^3\!\cos 3q_x.  
\end{eqnarray}

In first order perturbation theory in $J_1$ and $J_2$ the spectrum is
dispersionless on the Shastry-Sutherland line $J_1=2 J_2$ (also known
as 'disorder line'). Eq.~(2) also shows the general feature that the
location of the minimum of the dispersion curve shifts from $q_x=0$ to
$q_x=\pi$ somewhere close to crossing this line. 
 
In recent approximate theoretical treatments of interacting dimer
systems \cite{GopRS94,CHF99} the exchange interactions between
individual were reduced to an effective interaction between
dimers. For the present 1D zigzag chain the result of this
approximation is that only the combination $J_2 - \frac{1}{2}J_1$
enters into the dispersion. In more detail, the result is

\begin{equation} 
\omega(q_x) = \sqrt{1 + 2 \delta\omega^{(1)}(q_x)}
\end{equation} 
where $1 + \delta\omega^{(1)}(q_x)$ is the dispersion
in lowest order, i.e. the result of simple propagation of an excited
dimer triplet without considering its coupling to higher energy
modes. Evidently this is true in lowest order of the expansion; it is
seen, however, already from Eq.~(2) that additional terms which depend
on the individual exchange interactions enter in higher order.
Comparing the higher order coefficients in the series expansion we
find that the effective dimer approximation of Eq.~(3) amounts to
keeping only the leading (i.e. lowest) powers in $J_1, J_2$ for each
coefficient $a_n$ of $\cos(nq_x)$.
\begin{figure}
\mbox{\psfig{figure=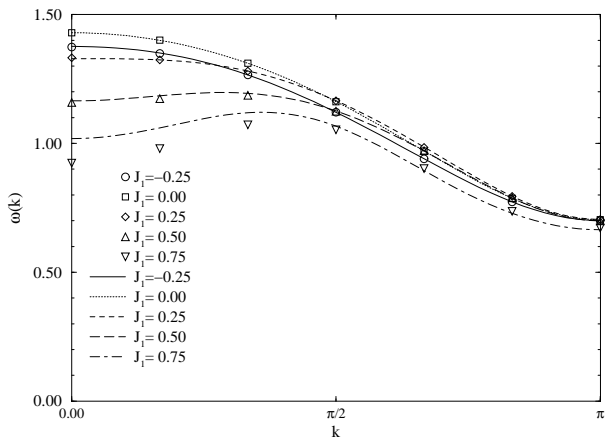,width=80mm,angle=-90.}}
  \vskip 2mm\nopagebreak
\caption{\label{fig:fig5mi} Comparison of excitation spectra for fixed 
effective dimer interaction $J_2 - \frac{1}{2}J_1 = \frac{3}{8}$.} 
\end{figure}
For a quantitative check of the effective dimer approximation we refer
first to ref. \onlinecite{SinZ99} where the dispersion on the disorder
line, i.e. for fixed $J_2 -\frac{1}{2}J_1 = 0$ is shown. The effective
dimer approximation is reasonable for a large part of the line but
deteriorates rapidly when the non-alternating limit $J_1 =1$, i.e. the
Majumdar-Ghosh point is approached. For the more generic value $J_2 -
\frac{1}{2}J_1 = \frac{3}{8}$ we show the dispersion in 
Fig.~\ref{fig:fig5mi}; it is
seen that the effective dimer approximation is of very limited value
in this case.
In addition these results, comparing the series expansion results
to spectra from exact diagonalization, we find the following limits 
of validity
for the dimer series expansion method: Whereas the $10^{\rm th}$ order 
spectra coincide with exact diagonalization results within a few
percent for $\vert J_1 \vert \le 1$ and $J_2 < 0.5$ the accuracy 
deteriorates rapidly when the symmetric ladder ladder configuration 
($J_1=0, J_2=1$) is approached.     

\begin{figure}
\mbox{\psfig{figure=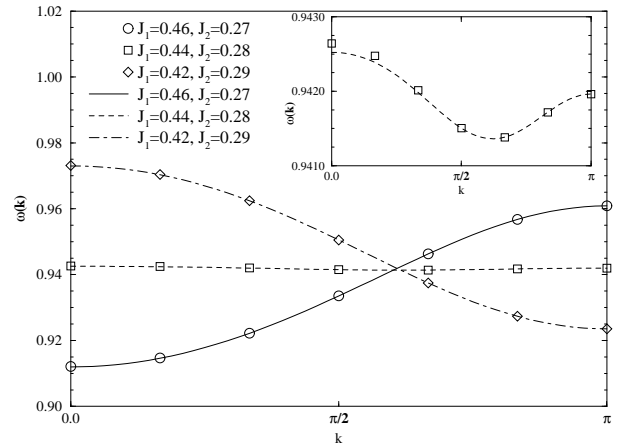,width=80mm,angle=-90.}}
  \vskip 2mm\nopagebreak
\caption{\label{fig:6a} Variation of the dispersion curves when the
incommensurate region is crossed. Inset shows with high resolution
an example of a dispersion curve with minimum at finite wavevector 
as obtained from exact (Lanczos) diagonalization as well as from the 
dimer expansion method.}
\end{figure}

\begin{figure}
\mbox{\psfig{figure=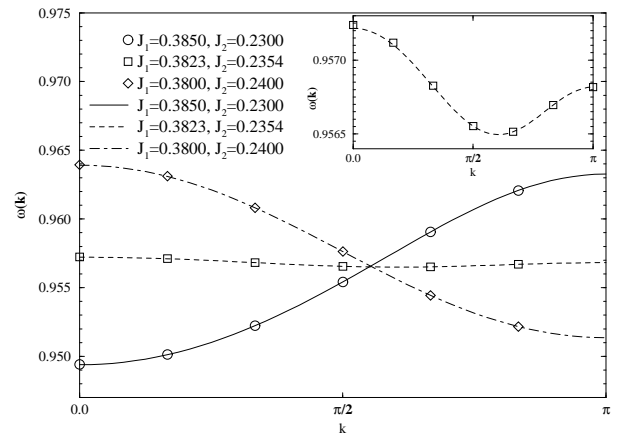,width=80mm,angle=-90.}}
  \vskip 2mm\nopagebreak
\caption{\label{fig:6} As Fig.~\ref{fig:6a} but for different couplings}
\end{figure}

\begin{figure}
\mbox{\psfig{figure=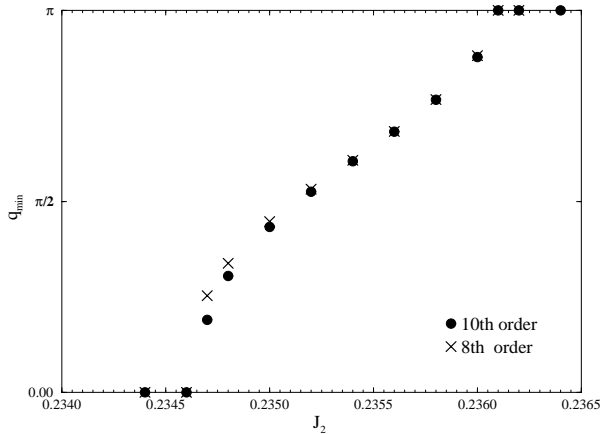,width=80mm,angle=-90.}}
  \vskip 2mm\nopagebreak
\caption{\label{fig:minicomp} Wavevector $q_{\rm min}$ for the minimum value 
of $\omega(q)$ on the line $J_1 = \frac{1}{2}(1 - J_2)$}
\end{figure}

Whereas the dispersion is flat on the disorder line to first order in
$J_1, J_2$, but develops maxima and minima in higher orders with an
apparent jump of the minimum energy form $q_x = 0$ to $q_x = \pi$
close to this line, it was noted already before \cite{BMN96,BKMN98}
from exact diagonalization results that a regime with extremely flat
dispersion and a minimum at finite wavevector $0 < q_x < \pi$ might
exist in a narrow regime in $J_1-J_2$ parameter space. We have
investigated this point again using the series expansions to $10^{\rm
th}$ order which provide us with a continuous wavevector dependence
and have confirmed the earlier speculation: In Fig.~\ref{fig:6a},
\ref{fig:6} we show two examples for spectra on the lines
$J_1 = 1 - 2J_2$ resp. $J_1 = \frac{1}{2}(1 - J_2)$: The
points in Fig.~\ref{fig:6a}
were discussed before in ref.~\cite{BMN96}; Fig.~\ref{fig:6a}
demonstrates that the shallow minimum of the dispersion curve at a
wavevector in the middle of the Brillouin zone is reproduced in the
series expansions. As further example we show in Fig.~\ref{fig:6} the
flat dispersion for somewhat smaller values of $J_1$. In Fig.~\ref{fig:minicomp} we show 
the development of the wavevector $q_{\rm min}$ for the minimum value of $\omega(q)$ on the line $J_1 = \frac{1}{2}(1 - J_2)$ in the $J_1-J_2$ parameter space (corresponding to the case of Fig.~\ref{fig:6}). The gradual transition of $q_{\rm min}$ between $q_{\rm min}=0$ and $q_{\rm min}=\pi$, possibly with infinite slope at the end points, is clearly seen. The comparison between results in $8^{\rm th}$ and $10^{\rm th}$ order illustrate the convergence of the expansion. At present,
however, the series expansions do not give a hint to a possible
fundamental reason for nearly flat dispersion: We have examined the
expansion coefficients $a_n$ of $\cos(nq_x)$ up to $10^{\rm th}$
order, but we do not find any indication of a convergence to zero for
$n \ne 0$.

\section{Interacting zig-zag Chains}

In real materials consisting of weakly interacting chains such as
${\rm KCuCl_3}$ \cite{KTTS98,CHF99} and ${\rm CuGeO_3}$ \cite{BR96} it
is clear from inelastic neutron scattering experiments that there is
considerable dispersion for wavevectors perpendicular to the double
chain direction. For weakly interacting chains, the series expansion
approach is to be considered as the only reliable systematic approach
which (by comparing results in subsequent orders) allows a consistency
check. Whereas the series expansions have been extended to cover
systems coupled in 2D and interesting results have been obtained 
for the spin Peierls material ${\rm CuGeO_3}$ \cite{Uhr97}, for the
1/5 depleted square material $\rm CaV_4O_9$ \cite{ZOH98} as well as
for general parameters \cite{KogK99}, we present in the following the
first results using the expansion for weakly interacting dimers for a
3D coupled system with particular application to investigate the
magnon dispersion in the material ${\rm KCuCl_3}$. The zig-zag chain
system ${\rm KCuCl_3}$ actually appears to be closer to the dimer
point than the systems treated so far and therefore is supposed to be a
better candidate for this expansion.

\begin{figure}
\mbox{\psfig{figure=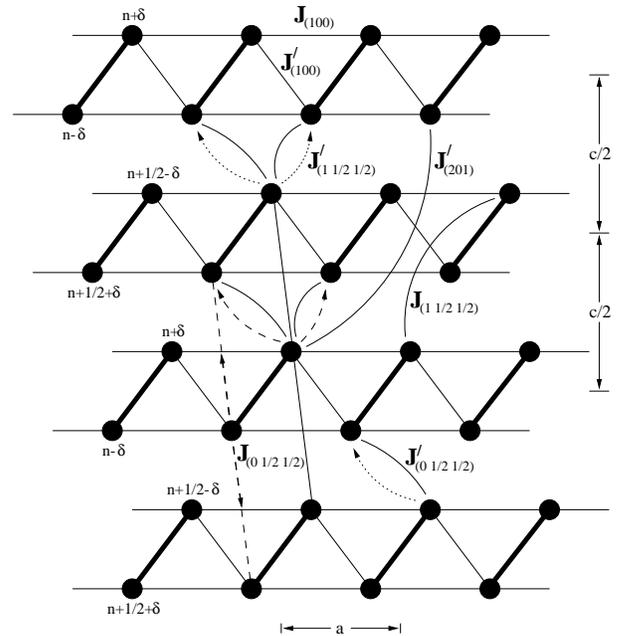,width=80mm,angle=0}}
  \vspace{3mm}
\caption{\label{fig:fig7} $\rm KCuCl_3-$structure projected on the
$xz-$plane.  Thick full lines denote the basic dimers, the height
above the reference plane $y=n b$ is given for the two spins of one
dimer in each zig-zag chain. Thin lines show interdimer interactions:
Full thin lines denote interactions of spins in dimers with identical
values of $n$, dashed and dotted thin lines denote interactions of
spins in dimers with different values of $n$. As indicated by the
arrows, the dashed (dotted) lines start at $n$ and end at $n- 1 \:
(n+1)$. Only one interaction of each type is shown.}
\end{figure}

The structure of ${\rm KCuCl_3}$ is shown schematically in Fig.~\ref{fig:fig7} in
a projection to what is conventionally called the $xz-$plane: The
fundamental dimers which are shown as solid lines form zig-zag chains
in the $x-$direction, neighbouring zig-zag chains are shifted with
respect to each other in $y-$direction by half a lattice constant
$\frac{b}{2}$; this shift as well as the tilting of the internal dimer
direction are indicated in Fig.~\ref{fig:fig7} by giving the $y-$coordinate for
each line of spins in $x-$direction ($n$ is an integer which numbers
the different planes). The elementary cell consists of two dimers,
dimer $D_1$ at the origin $\vec R_1=0$ and dimer $D_2$ at position $\vec
R_2 = (0,\frac{1}{2},\frac{1}{2})$ and the two spins forming each dimer, 1 and
2, are at positions $\vec R_i + \vec d_i$ for spin 1 and $\vec R_i -
\vec d_i$ for spin 2, for definiteness we take $d_{i,z} > 0$.

For the exchange interactions between spins we will use the following
notation:
The main intradimer exchange is denoted as $J$. The exchange
interaction per bond between spins in dimers separated by a lattice
vector $\vec R = la \vec e_x + mb \vec e_y + nc \vec e_z$ is denoted
as $J_{(lmn)}$ for the exchange between equivalent spins (pairs (11)
or (22)) of the corresponding dimers and as $J_{(lmn)}'$ for the
exchange between nonequivalent spins (pairs (12) or (21)). The
following exchange interactions will be considered: 
$J_{(100)}$ for pairs (11) and (22), $J_{(100)}'$ for the pair (12) and 
$J_{(201)}'$ for the pair (12), 
and for the two cases $p=0$ and $p=1$: 
$J_{(p\frac{1}{2}\frac{1}{2})}$ for the pair (11) starting from
dimer $D_1$  and for the pair (22) starting form dimer $D_2$ , 
$J_{(p-\frac{1}{2}\frac{1}{2})} = J_{(p\frac{1}{2}\frac{1}{2})}$ for
the pair (22) starting from dimer $D_1$  and for the pair (11)
starting from dimer $D_2$, 
$J_{(p-\frac{1}{2}\frac{1}{2})}' = J_{(p\frac{1}{2}\frac{1}{2})}'$ for
pairs (12) starting from either dimer $D_1$ or dimer $D_2$.

\begin{figure}
\mbox{\psfig{figure=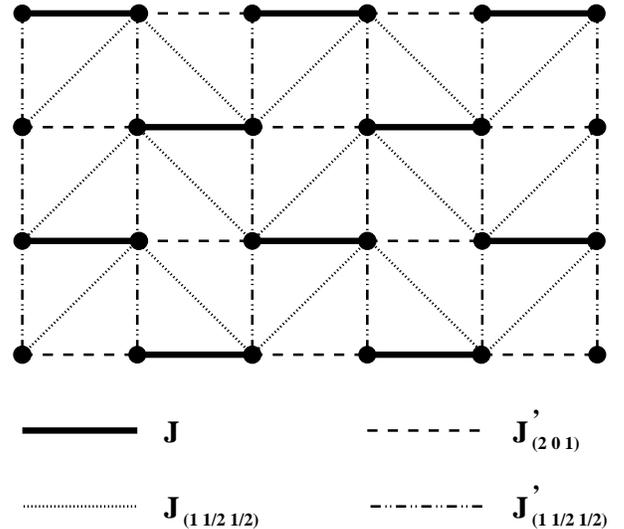,width=80mm,angle=-90.}}
  \vskip 2mm\nopagebreak
\caption{\label{fig:fig8mi} Alternative schematic view on the structure of 
$\rm KCuCl_3$.
}
\end{figure}

Alternatively we can look at the structure projecting on the plane
spanned by the directions $\vec e_y$ and $\vec e_x + \frac{1}{2} \vec
e_z$. In a schematic picture which shows the topology of the exchange
interactions only, we obtain Fig.~\ref{fig:fig8mi}; the 3D structure
of $\rm KCuCl_3$ results when identical planes are stacked and the
fundamental dimers are connected by zig-zag interactions. Evidently
the planar structure of Fig.~\ref{fig:fig8mi} can be reduced to a
number of limiting cases including coupled alternating chains
($J_{(1\pm\frac{1}{2}\frac{1}{2})} = 0$) or coupled ladders
($J_{(201)} = J_{(1\frac{1}{2}\frac{1}{2})}' = 0$).

If the dispersion is considered only to first order \cite{KTTS98} or
if higher orders are included in an RPA-like approximation
\cite{CHF99}, $J$ and $J'$ for a given $(lmn)$ enter only in the
combination $J^{\rm eff} = \frac 1 2 (pJ - p'J')$, where $p,p' = 1,2$ is the number of exchange
paths between the interacting dimers; $J^{\rm eff}$ is denoted as effective dimer interaction. 
As already seen in the 1D case of section 2, a correct treatment beyond first order involves $J$ and
$J'$ independently. According to previous work \cite{KTTS98,CHF99} the
main exchange interactions in addition to the basic intradimer
exchange are between dimers separated by $(lmn) = (100), (201),
(1\pm\frac{1}{2}\frac{1}{2})$. Except for $(201)$ these dimer-dimer
interactions involve both $J$ and $J'$ and it is our aim in the
following to discuss the validity of the effective dimer approximation
for $\rm KCuCl_3$ and to investigate to what extent the exchange
interactions between individual spins can be determined from the
present status of experimental results.

A calculation up to second order in the ratios $J_{(lmn)}/J$ leads to the
following expression for the frequency of the basic triplet (frequency
and coupling constants are measured in units of the basic dimer
exchange constant $J$ and wavevectors $q_i$ are given in units with 
the crystallographic lattice constants $a,b,c$ set equal to unity):
\begin{eqnarray}
&\omega(\vec q)&= 1 + \delta\omega^{(1)}(\vec q) -\frac 1 2 {\delta\omega^{(1)}}^2(\vec q) \nonumber\\
&+&J_{(100)}\!\left(J_{(100)}\!-\!J^{\prime}_{(100)}\right)-\frac 1 4 {J^{\prime}}^2_{(100)}\!\cos q_x\nonumber\\
&+&J_{\left(0 \frac 1 2 \frac 1 2\right)}J^{\prime}_{\left(0\frac 1 2 \frac 1 2\right)}+J_{\left(1\frac 1 2 \frac 1 2\right)}J^{\prime}_{\left(1\frac 1 2 \frac 1 2\right)} \nonumber\\
&+& \frac 1 2\left(J^2_{\left(0\frac 1 2 \frac 1 2\right)}\!-\!{J^{\prime}}^2_{\left(0\frac 1 2 \frac 1 2\right)}\right)\!\cos\!\frac{q_y} 2\!\cos\!\frac{q_z} 2 \nonumber\\
&+& \frac 1 2\left(J^2_{\left(1\frac 1 2 \frac 1 2\right)}\!-\!{J^{\prime}}^2_{\left(1\frac 1 2 \frac 1 2\right)}\right)\!\cos\!\frac{q_y} 2\!\cos\!\frac{2q_x\!+\!q_z} 2 \nonumber\\
&-&\frac 1 4 {J^{\prime}}^2_{(201)}\!\cos(2q_x+q_z)
\end{eqnarray}
Here
\begin{eqnarray}\label{eq:gl5}
&\delta\omega^{(1)}(\vec q)&= \frac 1 2 \left(2J_{(100)} - J^{\prime}_{(100)}\right)\cos q_x \nonumber\\
&+& \left(J_{\left(0\frac 1 2 \frac 1 2\right)}\!-\!J^{\prime}_{\left(0\frac 1 2 \frac 1 2\right)}\right)\!\cos\!\frac{q_y} 2\!\cos\!\frac{q_z} 2 \nonumber\\
&+& \left(J_{\left(1\frac 1 2 \frac 1 2\right)}\!-\!J^{\prime}_{\left(1\frac 1 2 \frac 1 2\right)}\right)\!\cos\!\frac{q_y} 2\!\cos\!\frac{2q_x\!+\!q_z} 2 \nonumber \\
&-&\frac 1 2 J^{\prime}_{\left(201\right)}\cos\left(2q_x+q_z\right)
\end{eqnarray}
is the dispersion to first order in the exchange constants, i.~e.\ the
effect of simple propagation of an excited dimer triplet.

In Eq.~\ref{eq:gl5} we have used for simplicity an extended zone scheme: Since
there are two dimers in the proper crystallographic unit cell, there
are two branches of triplet excitations for each wave vector in the
crystallographic Brillouin zone ($-\pi < q_yb, q_zc \le \pi$).  Owing
to the symmetry of the exchange interactions in the hamiltonian, these
two branches join to the unique smooth expression given in
Eq.~\ref{eq:gl5}, when the doubled zone $2\pi < q_zc + q_yb, q_zc - q_yb \le
2\pi$ is used. Each triplet excitation, however, is present at all
equivalent wavevectors of the crystallographic reciprocal lattice. For
a discussion of dipole transition amplitudes the noninteracting
triplet approximation can be used as sufficient guide and gives the
follwing results \cite{CHF99}: For momentum tranfer perpendicular
(parallel) to $\vec e_y$ the branch of Eq.~\ref{eq:gl5} in the first (second)
crystallographic Brillouin zone gives the only nonvanishing
contribution.  (A change form the first to second crystallographic
Brillouin zone then corresponds to a change in either $q_y$ or $q_z$ by
$2\pi$, i.e. to a change of sign in the second and third line of
Eq.~\ref{eq:gl5}).

We note that the wavevector $q_z$ should be distinguished 
from the quantity $q_{\perp}$ which is often used to denote the
two values of the quantum number parity discussed in section II
as $q_{\perp} = 0, \pi$. This wavevector is measured in units of
$\tilde{c}^{-1}$, where $\tilde{c}$ is the rung length, which 
differs from the reduced lattice constant $\frac{c}{2}$. In 
experiments so far only the basic triplet with $q_{\perp} = \pi$
has been observed and the interchange of minima with variation
of $q_z$ is an effect of the crystallographic lattice geometry and
not of the 1D ladder geometry. Excitations with $q_{\perp} = 0$
are excitations with two dimer quanta and have a minimum energy 
of twice the gap energy $\Delta$.

In order to obtain results which are quantitatively reliable we have
performed series expansions for the 3D coupled system to $4^{\rm th}$ order
following the lines described in section II. Because of the complex
lattice we did not characterize the clusters which results in a large
number of clusters: 5532 clusters in 4th order. The convergence of these series
expansion results is excellent for the small values of the expansion
parameters $J_{(lmn)}/J \leq 0.4$ in the application to $\rm KCuCl_3$.

\begin{figure}
\mbox{\psfig{figure=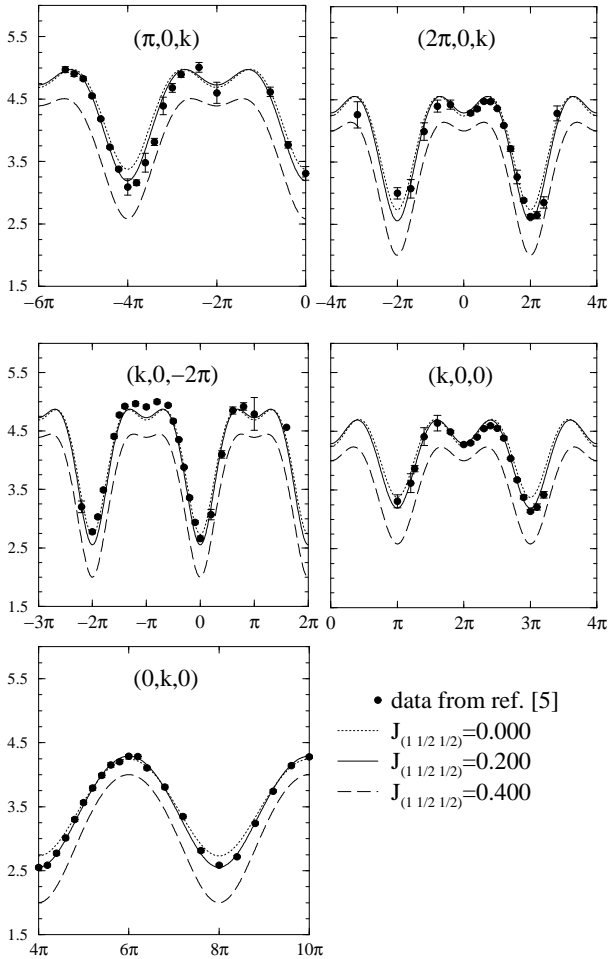,width=80mm,angle=0}}
  \vspace{3mm}
\caption{\label{fig9mi} Dispersion curves for $\rm KCuCl_3$ in various
directions in $\vec q-$space showing the variation with 
$J_{(1\frac{1}{2}\frac{1}{2})}'$ at fixed effective dimer interaction 
$J_{(1\frac{1}{2}\frac{1}{2})} - J_{(1\frac{1}{2}\frac{1}{2})}'=  0.160$} 
\end{figure}

\begin{figure}
\mbox{\psfig{figure=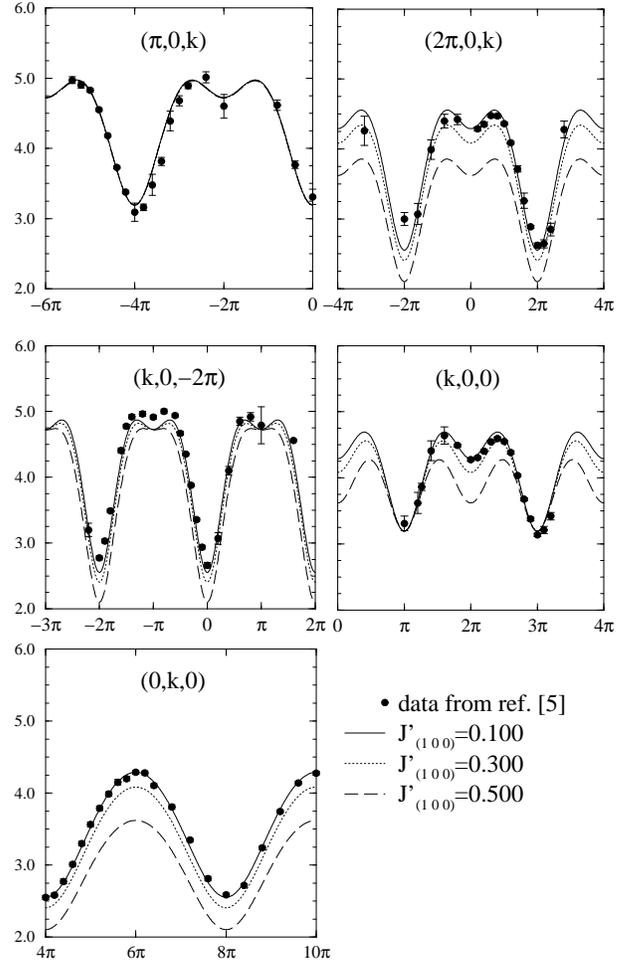,width=80mm,angle=0}}
  \vspace{3mm}
\caption{\label{fig10mi} Dispersion curves for $\rm KCuCl_3$ in various
directions in $\vec q-$space showing the variation with 
$J_{(100)}'$ at fixed effective dimer interaction 
$J_{100} - \frac{1}{2} J_{(100)}' = -0.055$}
\end{figure}

In Figs.~\ref{fig9mi} and ~\ref{fig10mi} we show results for
dispersions along typical lines in $\vec q-$space together with the data points
from ref.~\onlinecite{CHF99}.  Dispersions are plotted only 
for the branch  which has nonvanishing dipole transition amplitude in lowest order 
leading to $+$ or $-$ sign in Eq.~\ref{eq:gl5}
as discussed above. We assume that the effective dimer exchange takes 
the values determined in previous work
\cite{KTTS98,CHF99} and have therefore fixed the following
combinations of exchange parameters:

\begin{eqnarray}
J_{(201)}' &=& -2 J_{(201)}^{\rm eff} = 0.188, \nonumber\\
2J_{100} - J_{(100)}' &=& 2 J_{(100)}^{\rm eff} = -0.110,\nonumber \\
J_{(1\frac{1}{2}\frac{1}{2})} - J_{(1\frac{1}{2}\frac{1}{2})}'&=& 2 J_{(1 \frac{1}{2} \frac{1}{2})}^{\rm eff} = 0.160 
\end{eqnarray} 

The interactions $J_{(0,\frac{1}{2},\frac{1}{2})}, 
J_{(0,\frac{1}{2},\frac{1}{2})}'$ are found to be negligibly small.
In order to demonstrate the relevance of the individual spin exchange
parameters (as opposed to the effective description) we discuss the
separations $(100)$ and $(1\pm\frac{1}{2}\frac{1}{2})$ independently:
In Fig.~\ref{fig9mi} we present the variation of the dispersions 
$\omega(\vec q)$ in selected directions in $\vec q$ space
for different values of the coupling $J_{(1\frac{1}{2}\frac{1}{2})} = 0, 0.2, 0.4$ with the
remaining parameters fixed as given above. It is seen that the
distribution of the effective dimer interaction between parallel and
diagonal terms essentially shifts the dispersion curve by constant
amounts.  A comparison to the corresponding neutron scattering results
leads to the conclusion that $J_{(1\frac{1}{2}\frac{1}{2})} = 0.200$
is the most likely value. The analogous results for different values
of the diagonal (zig-zag) coupling $J_{(100)}' = 0.1, 0.3, 0.5$ are
shown in Figs.~\ref{fig10mi}. Here the frequency $\omega(q_x=0)$ (which is 
the energy gap for $q_z= \pi$ and the dip energy for $q_z = 0$) depends
only on the effective interaction, whereas the frequency
$\omega(q_x=\pi)$ (which is minimum for $q_z=0$ and a dip
energy for $q_z = 2\pi$) allows to determine the exchange between
individual spins.

Comparison of our series expansion results to the neutron scattering
data then leads to the following values for the exchange constants
between individual spins, not yet determined by the values for the
effective dimer interactions published so far:
\begin{eqnarray*}
J_{(100)}  &\approx & -0.005 \\
J_{(100)}' &\approx & \phantom{-}0.100.  
\end{eqnarray*}
These values imply that the ladder system in $\rm KCuCl_3$ is much
closer to an alternating spin chain than was believed so far and the leg
interaction tends to be ferromagnetic if it is nonzero at all. The
results for the interchain interactions in ($1 \frac{1}{2} \frac{1}{2}$)
direction are much less conclusive. The most likely values are 
\begin{eqnarray*}
J_{(1\frac{1}{2}\frac{1}{2})} &\approx& \phantom{-}0.200\\
J_{(1 \frac{1}{2}\frac{1}{2})}'& \approx& \phantom{-}0.040.  
\end{eqnarray*}
However, the error is large and data actually may be compatible also with
smaller values for $J_{(1 \frac{1}{2} \frac{1}{2})}$.

\section{Conclusions}

We have investigated the dispersion curve for the low energy triplet
excitations of one-dimensional and of weakly coupled zig-zag chains
starting from the limit of noninteracting dimers and performing an
expansion in the interdimer interactions. The series up to $10^{\rm th}$
order in the 1D case and up to $4^{\rm th}$ order in the 3D case were
evaluated explicitly after implementation on a work station. In the 1D
case the dispersion curves agree with those obtained from exact
diagonalization using the Lanczos algorithm for a large regime around
the dimer point; in this regime the method provides a reliable
approach to calculate the dispersion in its continuous dependence on
the wavevector. In a narrow regime close to the Shastry-Sutherland
line we find the minimum of the dispersion curve at an intermediate
wavevector $k_{min}, 0 < k_{min} < \pi$. As an application of the 3D
case we present the application of the method to the $\rm
KCuCl_3-$structure. Fitting to the dispersion as measured in inelastic
neutron scattering experiments \cite{CHF99} we determine the 
exchange interchange interactions between individual spins in
addition to the effective interaction between dimers determined 
before. It is shown that the effective dimer approximation, when treated
in the random phase approximation, sums up the leading powers
of the dimer expansion.  

\section*{Acknowledgement}
We gratefully acknowledge useful discussions with A. Kolezhuk and with
U. Neugebauer, who participated in the early stage of this work. We are
grateful to N. Cavadini for correspondance and for communicating the data points
of ref.~\onlinecite{CHF99} in detail. The work was supported by the German Ministry for 
Research and Technology (BMBF) under contract No. 03Mi5HAN5.

\end{document}